\documentclass{IEEEoj}
\usepackage{cite}
\usepackage{amsmath,amssymb,amsfonts}
\usepackage{algorithmic}
\usepackage{graphicx,color}
\usepackage{textcomp}

\def\BibTeX{{\rm B\kern-.05em{\sc i\kern-.025em b}\kern-.08em
    T\kern-.1667em\lower.7ex\hbox{E}\kern-.125emX}}
\def\OJlogo{\vspace{-10pt}\includegraphics[height=20pt]{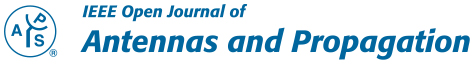}}

\usepackage{xcolor}
\usepackage{hyperref}

\begin{document}
\receiveddate{11 February 2025}
\reviseddate{14 March 2025}
\accepteddate{18 March 2025}
\publisheddate{24 March 2025}
\currentdate{28 May 2025. This preprint was compiled using IEEEoj.cls 2020/06/19 version V1.0}
\doiinfo{OJAP.2025.3554457}

\title{Beam Maps of the Canadian Hydrogen Intensity Mapping Experiment (CHIME) Measured with a Drone}


\author{
WILL TYNDALL$^{1,2,3}$,
ALEX REDA$^{1}$,
J. RICHARD SHAW$^{4}$,
KEVIN BANDURA$^{5,6}$, 
ARNAB CHAKRABORTY$^{2,3}$,
MARK HALPERN$^{4}$,
MAILE HARRIS$^{1,7}$,
EMILY KUHN$^{1,8}$, 
JOSHUA MacEACHERN$^{4}$,
JUAN MENA-PARRA$^{9,10}$,
LAURA NEWBURGH$^{1}$,
ANNA ORDOG$^{11,12}$,
TRISTAN PINSONNEAULT-MAROTTE$^{4}$,
ANNA ROSE POLISH$^{13}$,
BEN SALIWANCHIK$^{14}$,
PRANAV SANGHAVI$^{1}$,
SETH R. SIEGEL$^{2,3,15}$,
AUDREY WHITMER$^{1}$,
DALLAS WULF$^{2,3}$
}

\affil{Department of Physics, Yale University, New Haven, CT 06511 USA} 
\affil{Department of Physics, McGill University, Montreal, QC, Canada} 
\affil{Trottier Space Institute, McGill University, Montreal, QC, Canada} 
\affil{Department of Physics and Astronomy, University of British Columbia, Vancouver, BC Canada} 
\affil{Department of Computer Science and Electrical Engineering, West Virginia University, Morgantown, WV, USA} 
\affil{Center for Gravitational Waves and Cosmology, West Virginia University, Morgantown, WV, USA} 
\affil{Johns Hopkins University Applied Physics Laboratory, Laurel, MD, USA} 
\affil{Jet Propulsion Laboratory, California Institute of Technology, 4800 Oak Grove Drive, Pasadena, CA, 91109, USA} 
\affil{Dunlap Institute for Astronomy and Astrophysics, University of Toronto, Toronto, ON, Canada} 
\affil{David A. Dunlap Department of Astronomy and Astrophysics, University of Toronto, Toronto, ON, Canada} 
\affil{Department of Computer Science, Math, Physics, \& Statistics, University of British Columbia, Okanagan Campus, Kelowna, BC V1V
1V7, Canada} 
\affil{Dominion Radio Astrophysical Observatory, Herzberg Astronomy and Astrophysics Research Center, National Research Council Canada, Penticton, BC, Canada} 
\affil{Center for Astrophysics, Harvard and Smithsonian, Cambridge, MA 02138, USA} 
\affil{Instrumentation Division, Brookhaven National Laboratory, Upton, NY, 11973, USA} 
\affil{Perimeter Institute for Theoretical Physics, 31 Caroline Street N, Waterloo, ON N25 2YL, Canada} 

\corresp{CORRESPONDING AUTHOR: Will Tyndall (e-mail: william.tyndall@mcgill.ca).}

\authornote{IEEE OJAP encourages responsible authorship practices and the provision of information about the specific contribution of each author\\[1em]
This work was supported by the Natural Sciences and Engineering Research Council (NSERC) of Canada.}
\markboth{Preparation of Papers for \textsc{IEEE Open Journal of Antennas and Propagation}}{Author \textit{et al.}}




\begin{abstract}
We present beam measurements of the CHIME telescope using a radio calibration source deployed on a drone payload. During test flights, the pulsing calibration source and the telescope were synchronized to GPS time, enabling in-situ background subtraction for the full $N^2$ visibility matrix for one CHIME cylindrical reflector. We use the autocorrelation products to estimate the primary beam width and centroid location, and compare these quantities to solar transit measurements and holographic measurements where they overlap on the sky. We find that the drone, solar, and holography data have similar beam parameter evolution across frequency and both spatial coordinates. This paper presents the first drone-based beam measurement of a large cylindrical radio interferometer. Furthermore, the unique analysis and instrumentation described in this paper lays the foundation for near-field measurements of experiments like CHIME.
\end{abstract}

\begin{IEEEkeywords}
Antenna measurements, telescopes, unmanned aerial vehicle (UAV), unmanned aerial system (UAS)
\end{IEEEkeywords}


\maketitle


\section{INTRODUCTION}
Future and current 21\,cm intensity mapping telescopes such as CHIME \cite{CHIMEoverview}, CHORD \cite{CHORD_Instrument}, HERA \cite{HERAoverview}, HIRAX \cite{2022JATIS...8a1019C,HIRAX_Instrument}, LOFAR \cite{LOFARoverview}, MWA \cite{MWAoverview}, SKA \cite{SKAoverview} \cite{Meerkat} will measure the distribution of neutral hydrogen across a wide range of redshifts for a variety of science goals \cite{CV_Dark_Energy}. Isolating the cosmological 21\,cm signal by removing bright radio foregrounds \cite{Shaw_polmodes, tcmreionizationforegrounds, foregroundsubtractrequirementstcm, parsonsdelayspecfilterforeground, thyagarajanforegroundswidefieldspectra, dayenuforegroundfilter.500.5195E} will require precision beam measurements to better than 1\% uncertainty \cite{shaver1999, oh2003, liu2011, Shaw_polmodes, CC_drone_mapping}. Calibration with well known radio point sources is difficult for stationary (un-steerable) drift scanning instruments, and one promising alternative is to use radio sources onboard drones as piloted calibration references \cite{CC_drone_mapping, JacobsDroneECHO}.

Decades of technical innovations and cost reductions have improved the feasibility of using unmanned aerial vehicles (UAVs) as precision instruments in a research setting, particularly for calibrating microwave and radio telescopes and antennas \cite{Virone2014NF, Pupillo2015, GarciaFernandez2017, Fritzel2016, CulottaLopez2021, Zhang2021, Bolli2018}. An additional benefit of using drones is the capability of measuring antenna performance on site, within the configuration of the larger array and surrounding environment \cite {JacobsDroneECHO, MauermayerNFdrone}. With high precision positional accuracy, autopilot software, and carefully designed flightplans, UAV calibration can directly measure the far-field beam pattern of telescopes and arrays. Where the far-field is inaccessible the (complex) near-field electric field could be measured and algorithmically transformed into far-field beam patterns \cite{Virone2014NF, MauermayerNFdrone, CC_drone_mapping, Parini2022, GarciaFernandez2017}.

CHIME is a radio interferometer operating between 400-800\,MHz to measure redshifted neutral hydrogen (see \cite{CHIMEoverview} for more detail). The cylindrical reflectors observe the sky by drift scanning with its large field of view. The resulting primary beam poses a unique challenge for drone beam mapping for several reasons: (1) the far-field is distant, at altitudes in excess of 1\,km at all frequencies, (2) it is highly elliptical (1.2-2$^{\circ}$ East-West, and $\sim 120^\circ$ North-South) with frequency- and declination-dependent variations \cite{dallassolar,CHIMEoverview,Redaholography}.

In this paper we present measurements of the CHIME primary beam obtained with drone-based beam mapping techniques. We acquire these measurements in a region which overlaps with existing solar \cite{dallassolar} and holography \cite{galt_spie_proc,Redaholography} beam measurements. To detect the pulsing drone calibration signal we utilize the pulsar gating mode of CHIME to perform in-situ background subtraction. The resulting comparisons we present here provide independent validation of sky-based beam models and demonstrate the synchronization of the calibrator and receiver enabling future measurements of CHIME with a drone-based platform (e.g. polarization, repeatability, near-field to far-field). 

Due to flight altitude restrictions, the flight took place squarely in the radiating near-field, hindering direct comparisons of the beam as a function of angle between drone data, solar data and holography data sets. Instead, we compare best-fit main beam parameters and other beam properties across frequency and declination to test the feasibility of these methods. These measurements will inform future near-field beam mapping efforts and near-field to far-field transformations of measured CHIME beams. 

In Section \ref{sec:methodology} we discuss the methodology used in this paper, including the CHIME telescope and drone instruments, the CHIME configuration developed for drone beam mapping, and the list of performed drone flights. In Section \ref{sec:DR} we describe the data sets and data processing we developed to compare drone beam measurements to solar and holography beam measurements, as well as an independent drone altitude determination. In Section \ref{sec:results} we present comparisons between the drone and solar data maps and fitted primary beam quantities across feed, frequency, and spatial coordinates. 


\section{METHODOLOGY}
\label{sec:methodology}
\subsection{CHIME} 
The Canadian Hydrogen Intensity Mapping Experiment (CHIME) is a radio interferometer observing from the Dominion Radio Astrophysical Observatory (DRAO) in British Columbia, Canada, whose array center is located at longitude 49.3208N, latitude -119.6236E and height $\sim$545m above sea level. CHIME's primary science objective is to measure the distribution of neutral hydrogen (HI) over the sky as a biased tracer of the underlying distribution of matter \cite{Pen:2009, Chang:2010, Masui:2013, Switzer:2013, Anderson:2018, Wolz:2021, Shaw_polmodes, CHIMEdetection, CHIMELymanDetection}. The statistics of this distribution probe a cosmological standard ruler, the baryon acoustic oscillation (BAO) scale. Tracking the evolution of this scale over cosmic time (i.e. redshift) in turn provides a constraint on the expansion history of the Universe (see most recent results in \cite{DESI_arxiv} and references therein).

To this end, CHIME is designed to observe the redshifted 21\,cm emission of HI over a wide band between 400-800\,MHz, corresponding to redshift $z = 0.8-2.5$, considered a critical epoch for constraining competing models of dark energy vis-a-vis their impact on the expansion history. The CHIME instrument is described in more detail in \cite{CHIMEoverview}, briefly, the telescope consists of four parabolic cylinders, 20\,m wide (East-West) by 100\,m long (North-South). Each cylinder is outfitted with 256 dual-polarization cloverleaf antennas separated by $\sim$30\,cm within the central 80\,m length of the focal line. The telescope has no moving parts; the cylindrical reflectors provide a beam shape which is focused in the transverse East-West direction but extends nearly from horizon-to-horizon in the North-South direction. This accommodates a ``driftscan" observing strategy; every 24 hours the telescope observes the entire sky accessible from the CHIME latitude due to the Earth's rotation. This stationary design complicates beam calibration, as we cannot ``point" the beam at a suitable calibrator. Alternatively, in this paper, we deploy a remotely controlled drone outfitted with a radiating payload as calibrator source to sample the instrument field of view.

\subsection{Drone Instrumentation}
Flights are performed with a payload mounted on a DJI Matrice 600 Pro with a Real-Time Kinematic (RTK) GPS accessory to provide $\sim$1\,cm location accuracy. The payload consists of a broad-band white noise generator, a bandpass filter that passes signals between 400-800\,MHz\footnote{\hyperlink{https://www.minicircuits.com/pdfs/ZABP-598-S+.pdf}{Minicircuits ZABP-598-S+ bandpass filter}}, and a switching board which either passes the white noise signal or terminates on 50$\Omega$ on a trigger. The trigger is generated using a separate GPS clock in the payload. For measurements on CHIME, the signal was set to pulse on at the start of a pulse-per-second (PPS) trigger from the GPS, remain on for 0.5\,s, then turn off for 0.5\,s, generating a sequence synchronized to the GPS pulse-per-second. 

The signal is transmitted from a single-polarization biconical antenna\footnote{\hyperlink{https://aaronia.com/en/produkte/antennas/bicolog}{Aaronia Bicolog 30100}}, the same used in \cite{JacobsDroneECHO}, with a frequency range operating between 30MHz-1GHz. A commercial bandpass filter was included in the analog chain to restrict the transmission signal to 400-800\,MHz. The transmitting antenna is isolated from the drone and the payload by a perforated aluminum groundplane. Although the carbon fiber legs and struts are reflective across our frequency band, they retract shortly after takeoff and remain stowed behind the groundplane during flight. Range measurements of the assembled drone system (consisting of the drone, payload, groundplane, and transmitting antenna) indicate that the transmitting antenna has an omnidirectional shape, as expected, with full-width-half-max (FWHM) values of $60^{\circ}$ in the frequency ranges presented here. All measurements presented are within $3^{\circ}$ of the antenna pointing, and thus the effect of the transmission beam is expected to be negligible. As a result, we do not account for the transmission beam pattern here. The range measurements also indicated that the cross-pol from the transmitting antenna is typically -20dB or better and is featureless within the inner few degrees. We therefore expect that in the region where CHIME data is presented, the beam and polarization properties of the transmitting antenna can be assumed to be isotropic.

Our measurement is equivalent to a Ludwig-I polarization measurement instead of the more common Ludwig-III measurement in spherical coordinates \cite{Ludwig}. This is a combination of two choices: first, the linear East-West flight path (described in more detail in Section~\ref{sec:flights}) means that the drone did not maintain a constant radial distance from the center of CHIME. Second, the omnidirectional antenna is attached to the drone body without a gimbal mount and so the transmitted polarization axis is fixed to the drone body and thus the polarization angle remains parallel to CHIME during the course of the measurement. Although more complex flight paths and gimbal mounts\cite{gimbal} are an interesting direction for future measurements, the data presented here is restricted to the inner few degrees of the beam where these projection effects are $<$1\%. Most importantly, our measurements are in the near field without a transformation to the far-field, such that we do not present drone measurements with the interpretation that they are far-field CHIME beams.

\begin{figure}
\centerline{\includegraphics[width=3.5in]{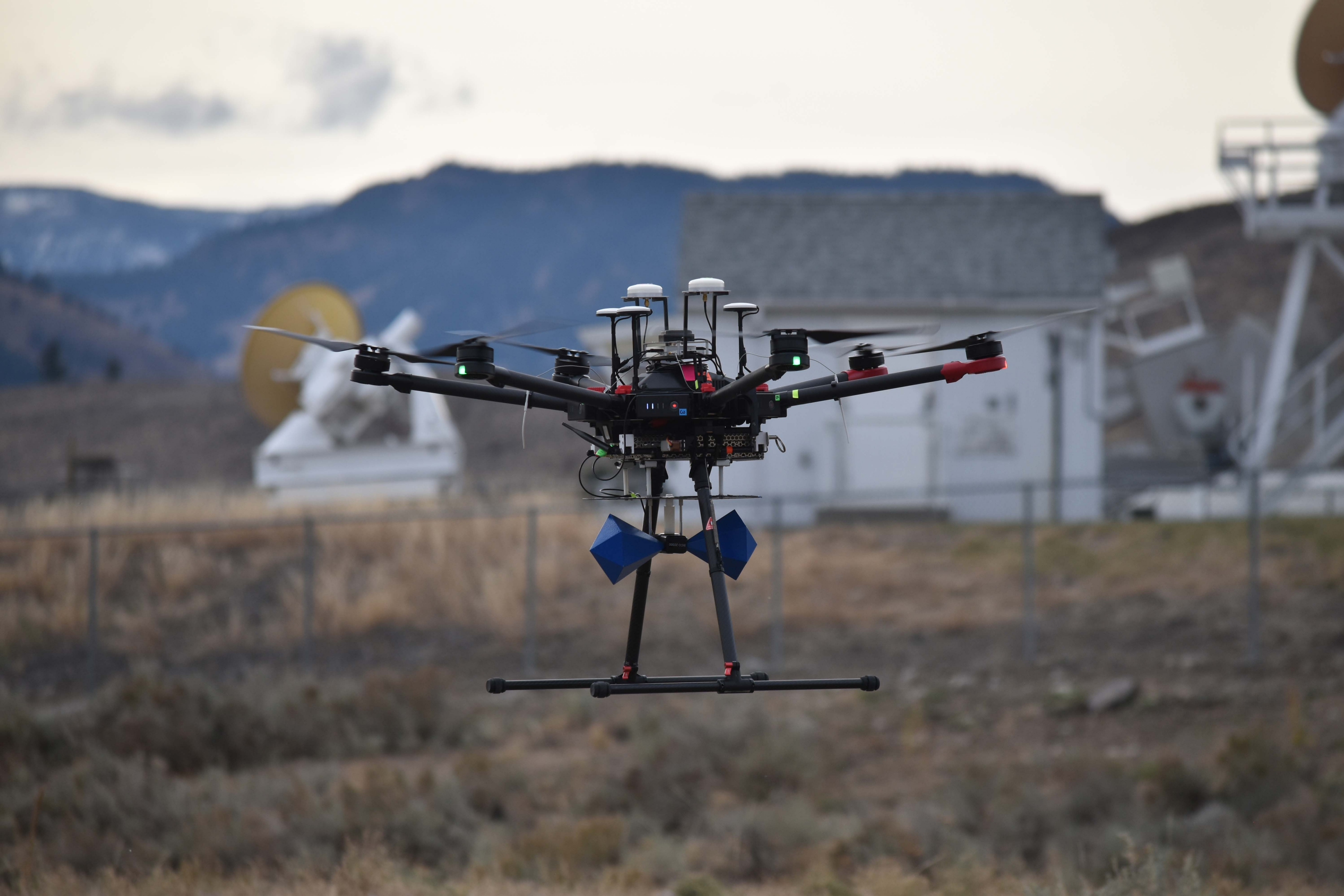}}
\caption{Photograph of the Matrice 600 Pro drone with payload, in flight at the DRAO during measurements.}
\label{dronepic}
\end{figure}

\subsection{CHIME Configuration for Drone Beammapping}
\label{sec:config}

As described in \cite{CHIMEoverview}, CHIME is an interferometer with 1024 dual-polarized antennas distributed evenly across four cylindrical reflectors. The raw data from each of the $N=2048$ inputs is digitized and channelized into 1024 frequency bins between 400-800\,MHz in an ICEBoard\cite{ICEboard}. Signals from each pair of antennas are then correlated together in a GPU correlator, forming a data set of visibilities with 1024 frequency bins and $N^{2}$ ($\sim$4x$10^{6}$) channels, typically integrated to a cadence of 10\,s. As we describe below, this data acquisition system required modifications to allow a measurement of the drone signal.

The flashing drone calibration source is seen by all CHIME feeds, and the on- and off-times are differenced and then accumulated as an $N^{2}$ matrix of visibilities using the pulsar gating mode available from CHIME. To avoid smearing within an integration frame we must reduce the accumulation time from the standard 10\,s down to 2\,s, but this increased data rate of around 50\,TB/hour can neither be offloaded from the CHIME X-engine nor can any useful length of data be stored within the on-site storage. To get around this we used the existing capabilities of the CHIME Science Data Processor (SDP) system to achieve a significant compression of the data\cite{CHIMEoverview}.

For a source seen by all feeds, the resulting $N^{2}$ visibility matrix is effectively rank-1, meaning the majority of the signal power is captured in a single common mode. This common mode can be identified by performing a single-value decomposition (SVD \footnote{\href{https://numpy.org/doc/stable/reference/generated/numpy.linalg.svd.html}{numpy.linalg.svd documentation}}) on the matrix such that the common mode will be the first eigenmode of the matrix. Although any motion of the drone within an integration frame can generate additional modes, we found at 2\,s cadence it is sufficient to record only a small number of the eigenvalues and vectors of the matrix to capture all of the information. Generating this eigendecomposition is computationally costly, and CHIME was designed to be able to do this once per 10\,s frame for its source-based gain calibration. To fit within the computational footprint we only include data from a single cylinder of the instrument, which further reduces the output data rate. As we use an iterative algorithm for this calculation\cite{CHIMEoverview} this gives only a factor of 16 computational savings, rather than the factor of 64 savings we would expect for a full calculation.

The set of modified processing stages we perform on the $N^{2}$ matrix, per frequency, are:
\begin{itemize}
    \item Use the existing CHIME pulsar gating capability to take on-off differences of the calibration source signal on the full $N^2$ visibility matrix. This is possible as both CHIME's clock and that of the noise source are locked to the GPS pulse-per-second (PPS) signal. This is then integrated to 2\,s cadence.
    \item The submatrix corresponding to the feeds within cylinder C (the cylinder to the east of the array center, see Figure~\ref{CHIMEcoordinates}) is extracted.
    \item The existing SVD capability is used to calculate the largest 10 eigenvalues and their eigenvectors \cite{CHIMEoverview}. 
\end{itemize}
These eigenmodes, as well as the full visibility matrix for cylinder C, is then sent to the receiver node of the SDP system for the final processing, and writing of the following data products: 
\begin{itemize}
    \item The full $N^{2}$ visibility matrix is written out for a set of 64 frequencies to allow us to verify that the eigendecomposition extracted all of the information.
    \item For the full set of 1024 frequencies we write out the eigenvalues and eigenvectors to allow full sensitivity analysis of the data, as well as measurements of phase variation.
    \item For the full set of frequencies we write out the 512 autocorrelations from both polarizations of the cylinder C feeds. This gives a simple and compact dataset, but at lower sensitivity and without the ability to constrain phase variations.
\end{itemize}
The gated power for each feed can be reconstructed using the strongest 10 eigenmodes from a single cylinder. We found the ratio of the first eigenmode to the second mode was 100 or higher within the main beam, indicating that the majority ($>99\%$) of the signal was the common mode response to the drone signal. As a result, the analysis in this paper uses the first eigenmode only, which sufficiently captures all of the drone signal while allowing for a lower noise floor than what is possible using the directly measured power contained in the autocorrelations.

\subsection{Drone Flights} 
\label{sec:flights}
We performed ten flights over the CHIME telescope; in this paper we present results from the flight at the approximate declination of Virgo A, where independent beam measurements of CHIME from both holography data \cite{galt_spie_proc} and solar data \cite{dallassolar} are available. Replicating the trajectory of Virgo A and the Sun through the CHIME primary beam required a 35$^{\circ}$ zenith angle drone flight relative to the array center. Due to altitude restrictions, we performed a low velocity flight at a conservative altitude to maximize the spatial resolution of the transit data. As shown in Figure~\ref{CHIMEcoordinates}, this flight was performed $\sim$220\,m South from the center of the array, corresponding to distances between 180 - 250\,m with respect to feed location along the focal line. Relative to the CHIME array center, the flight spanned -100\,m to +100\,m from East-to-West at an altitude of $\sim$307\,m, in accordance with our flight ceiling limits from TransportCanada. The first flight was flown with a heading of $\sim0^{\circ}$ [North] to measure co-polarizations corresponding to CHIME's N (YY) polarization, and the second flight was flown with a heading of $\sim90^{\circ}$ [East] to measure the orthogonal (XX) polarization. The typical distance from CHIME for these flights ranges from 350\,m - 390\,m. As a result, this flight plan corresponds to the near-field for all CHIME antennas (the CHIME far-field for a single feed is $>$1\,km). The time for each polarization pass was 3.4\,mins.


\section{DATA PROCESSING}
\label{sec:DR}

\begin{figure}
\centerline{\includegraphics[width=3.5in]{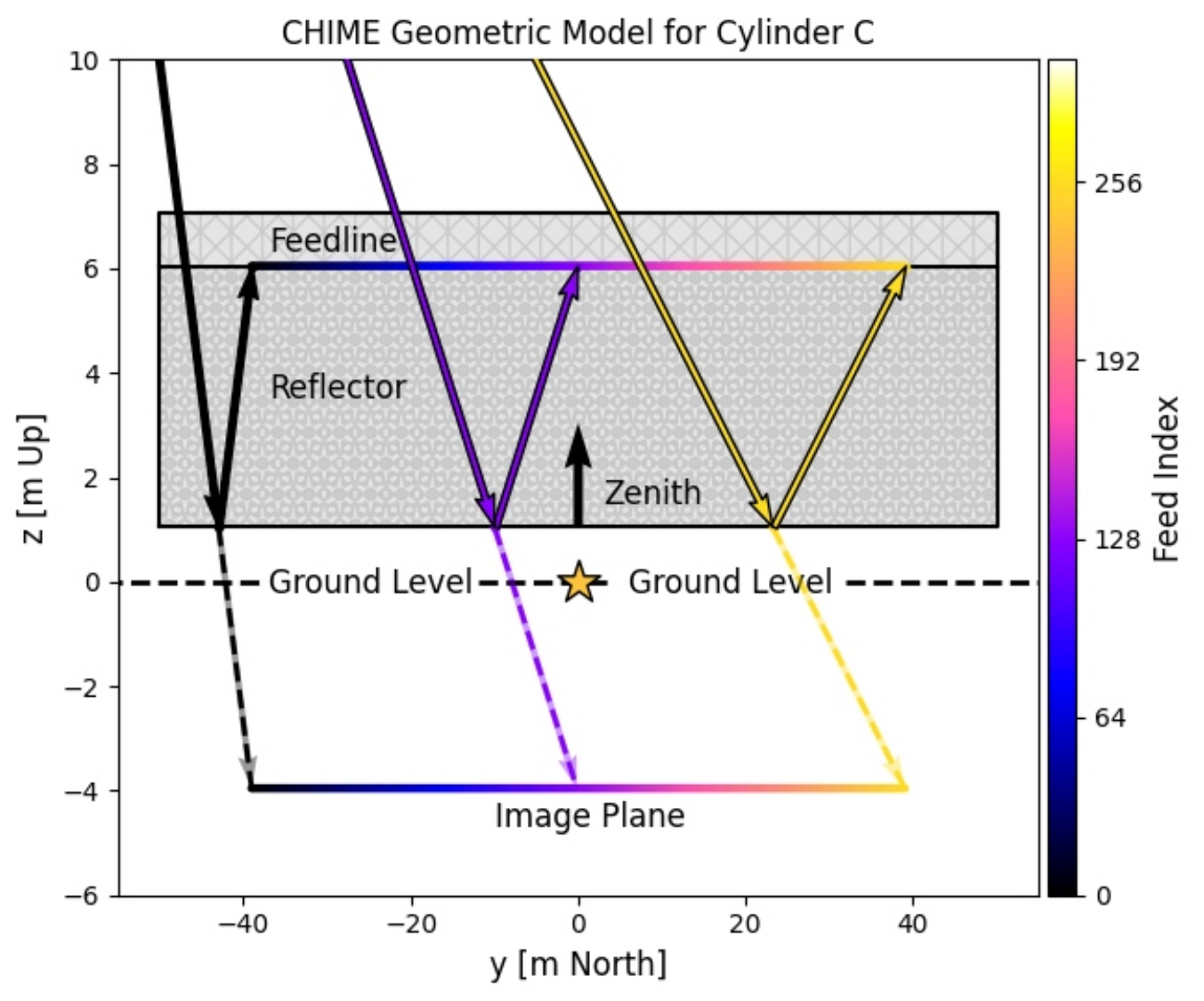}}
\caption{Optical path of calibration signal from drone-based transmitter as seen by CHIME feeds. The incoming radio waves are focused by the cylindrical reflector onto the feeds, which each observe the drone at a different elevation. In the reference frame of the southernmost feed (Feed 0, black) the drone appears to be higher in the sky than for feeds in the middle (Feed 127, purple) or at the North (Feed 256, yellow) end of the cylinder. The drone altitude in this figure is substantially lower than the flight altitude to exaggerate the feed-dependent parallax. The reference frame of the local Cartesian coordinate system is the survey location of the center of the CHIME array (yellow star). See text for more details. \label{CHIMEray} }
\end{figure}

\begin{figure}
\centerline{\includegraphics[width=3.5in]{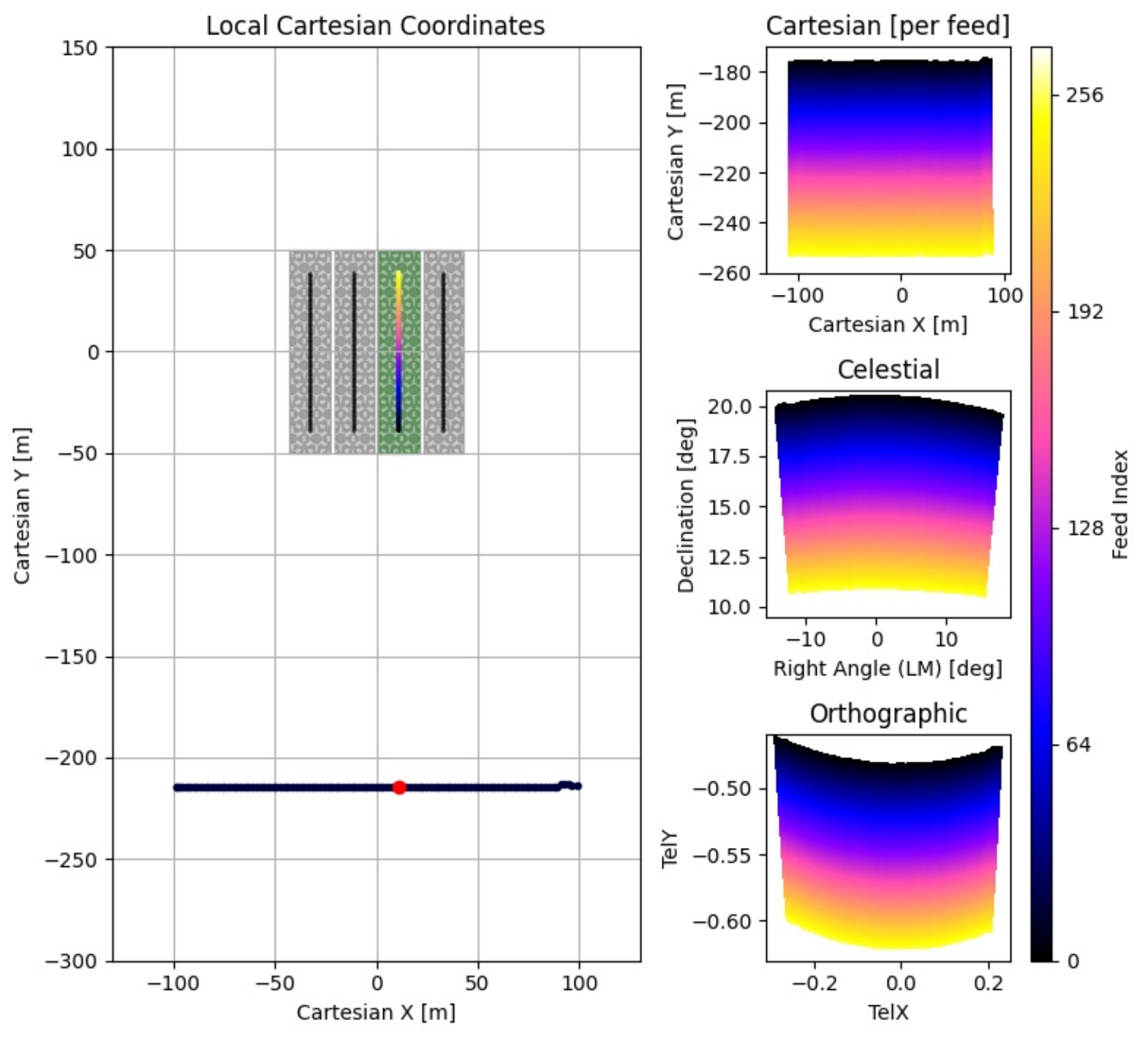}}
\centerline{\includegraphics[width=3.5in]{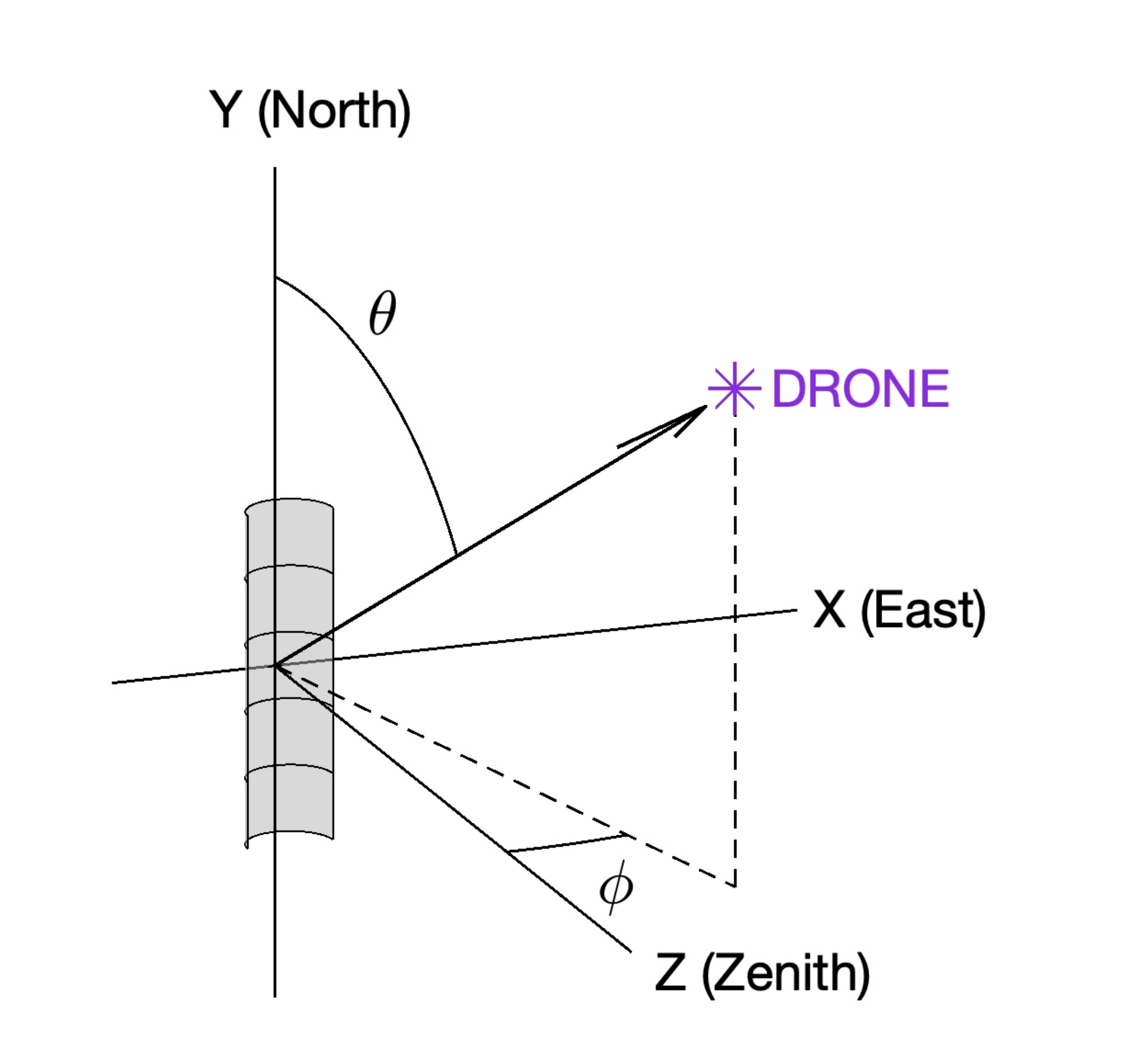}}
\caption{The upper half of this figure shows coordinate conventions for the data from a drone flight at 35 degrees of zenith angle. One transit (upper left)  across the beam of Cylinder C is shown true to scale in a local Cartesian coordinate system centered on the CHIME telescope. This flight trajectory is represented in three per-feed coordinate systems: Cartesian (upper right), Celestial (middle right) referenced to Local Meridian, and Orthographic (TelX/TelY)(lower right) using the color scale from Figure~\ref{CHIMEray}. The spherical coordinate system (bottom) with axes and angle definitions given to show the transformation from a spherical coordinate system to TelX (projection onto X-axis) and TelY (projection onto Y-axis).
\label{CHIMEcoordinates}}
\end{figure}

\subsection{Map Projections}
\label{sec:mapproj}
During data processing, each CHIME correlation frame is associated with a concurrent drone position measurement from the RTK unit, which expresses the drone location in latitude, longitude, and height coordinates. A conversion to a local Cartesian reference frame\footnote{\href{https://pypi.org/project/PyGeodesy/}{PyGeodesy}} is performed using the latitude, longitude, and height of a known reference point (in this instance, the center of the CHIME array provided by surveying). In order to compare our measurements with other CHIME data sets, we must project our coordinates into a compatible reference frame, in particular, we choose TelX/TelY (described in more detail below) because it allowed us to compare to solar data results~\cite{dallassolar}.

The drone flew a trajectory from East to West and then West to East, about 220\,m from the center of CHIME, as shown in local Cartesian coordinates in Figure~\ref{CHIMEcoordinates}, with Cylinder C highlighted green.  As shown in Figure~\ref{CHIMEray}, when the drone is directly on the boresight of CHIME cylinder, each feed will `see' the drone at a slightly different angle up from the horizon (or down from zenith). This angle crucially depends on the altitude of the drone flight. The altitude must be measured relative to the phase center of the array (where incident rays are focused onto the feedline). We model the phase center of each feed by extending the incident ray to an \textit{image} of the feedline placed beneath the primary reflector at a distance of 5\,m (the focal length). This defines the image plane, also shown in Figure~\ref{CHIMEray}. The height of the CHIME cylinders above ground is $\sim$1\,m, thus the image plane sits $\sim$4\,m below the ground surface. As a result, if the drone reports a height of 307\,m, we would expect to use 312\,m (above the image plane) to find the correct angle. We will independently constrain the effective altitude in Section~\ref{sec:DR}).

Once this drone altitude is established, we convert the positions surveyed during the drone flight from per-feed local Cartesian coordinates to TelX/TelY coordinates. The mapping of the drone flight path from local Cartesian to both celestial coordinates and TelX/TelY are shown in the right three panels of Figure~\ref{CHIMEcoordinates}, where the colorbar indicates feed number. 
The transformation from a spherical coordinate system to TelX/TelY is shown in the lower panel of Figure~\ref{CHIMEcoordinates}. In this coordinate system, the X-axis points East towards the horizon, and the Y-axis points North towards the horizon, and zenith (looking up) is defined along the Z-axis. TelX is the projection onto the X-axis, and represents the CHIME beam pattern parallel to the E-W axis; TelY is the projection onto the Y-axis, and represents the CHIME beam pattern parallel to the N-S axis. The resulting transformation is defined as:

$$
\mathrm{TelX} = \sin(\theta)\sin(\phi) 
$$
$$
\mathrm{TelY} = \cos(\theta)
$$

Where $\theta$ is the angle from North and $\phi$ is the angle from Zenith measured in the XZ-plane. In this system, celestial sources will follow a curved path in TelX/TelY as can be seen in Figure~\ref{overlap}. The choice of coordinate system was described in detail in \cite{dallassolar}, which presents solar data projected into the orthographic coordinate system (TelX, TelY). The advantage of this projection is that it orthogonalizes the features of the beam, and thus we use this projection throughout this paper.
 
\subsection{Available Datasets}
\begin{figure}
\centerline{\includegraphics[width=3.5in]{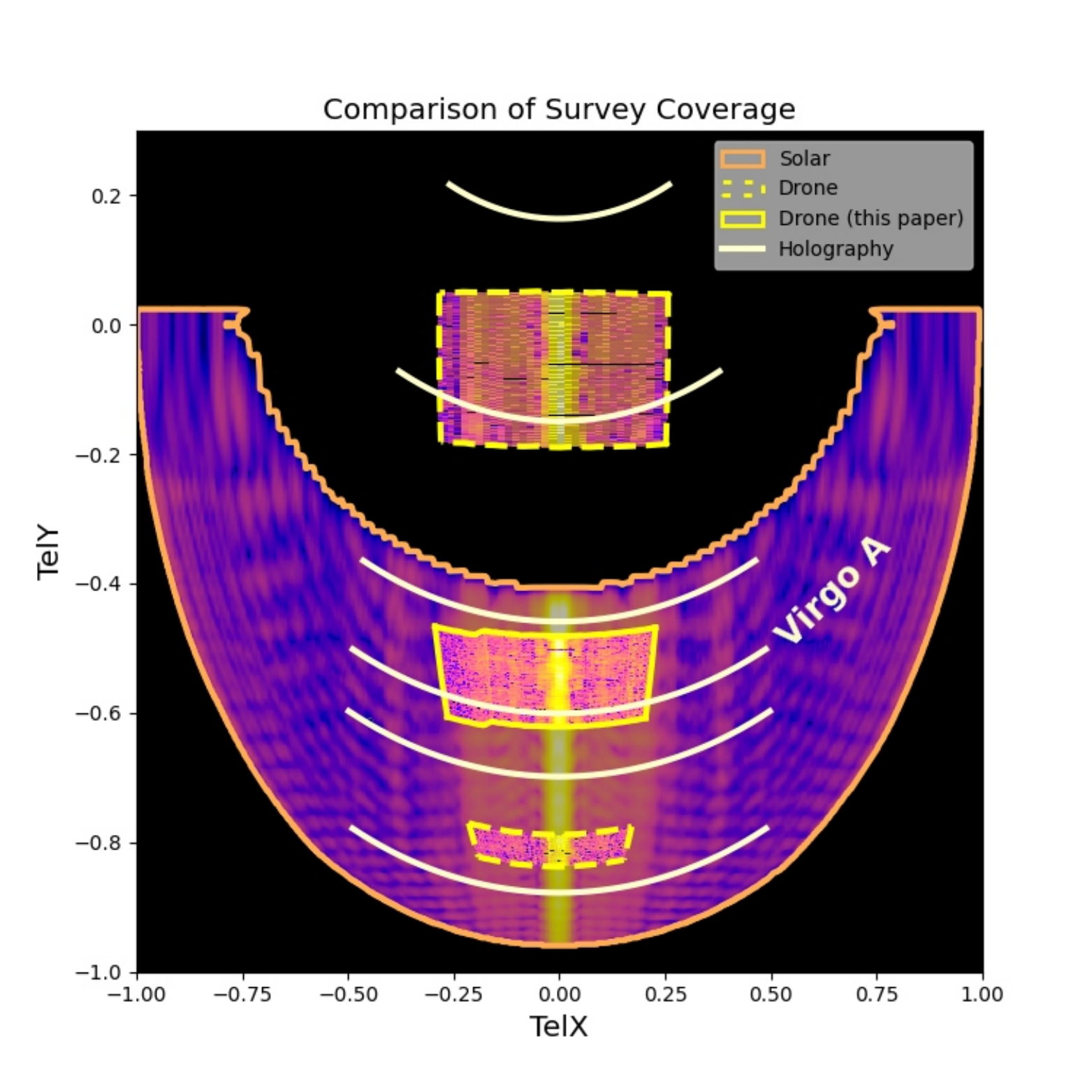}}
\caption{Beam characterization measurements have been calculated from three unique data sets acquired in different portions of CHIME's field of view. Coordinate axes are TelX and TelY, explained in more detail in Figure~\ref{CHIMEcoordinates}, which correspond to the projection to East and North, respectively. Solar beam measurements are presented in a continuous semicircular slice. Holographic measurements of galactic radio sources appear as pale yellow arcs. Measurements from the drone trajectory shown in Figure~\ref{CHIMEcoordinates} are indicated by the square patch with solid yellow outline. Some additional drone flights (not analyzed here) are indicated by patches with dashed yellow outlines. The color scale of the solar and drone datasets is proportional to received power to illustrate the alignment of the main beam, but has not been appropriately calibrated for direct comparison. Comparison of the FWHM and Centroid measurements obtained from these data sets are shown in Figures \ref{CHIMEFWHM} \& \ref{CHIMEcent}. \label{overlap}}
\end{figure}

Comparisons are possible between three independent data sets (a) Drone measurements; (b) Solar measurements; and (c) Holography measurements. The overlap between these data sources can be seen in TelX/TelY coordinates in Figure~\ref{overlap} and given in Table~\ref{tab:surveyparams}. In the drone flights described above, because each feed will map to a different angle, the single East-West flight maps to a range of TelY values. As can be seen, the drone data overlaps the solar transit data and with one source measured via holography, Virgo A. All three data sets overlap at TelY $\sim-0.6$ which corresponds to a declination of $\sim 12^{\circ}$ in astronomical coordinates.

The drone data presented here are the first eigenmode (from the decomposition described in Section \ref{sec:config}) for each feed on Cylinder C, and are measured in the near-field of CHIME. As described in\cite{dallassolar}, the solar data products are averages of cross-correlation visibilities of feeds within 10\,m of each other, restricted to pairs of feeds within the same cylinder (intracylinder cross-correlations with baseline distances $\le$10\,m). 
These short baselines retain the solar signal, allowing an estimate for the CHIME primary beam averaged across all feeds and cylinders. The holography data\cite{Redaholography} directly measures the far-field primary beam of each CHIME feed, and so feeds in Cylinder C can be selected and compared with the drone data. The holography data can also be processed with the same averaging methods used for the solar data for more direct comparisons across those two datasets. The parameter space spanned by these three data sets is given in Table \ref{tab:surveyparams}.

Although the choice of TelX/TelY orthogonalizes the beam shape to align with the geometry of CHIME, sources at constant declination such as the holographic measurements of Virgo A and each day of solar data are curved in TelX/TelY (see Figure \ref{overlap}), which distorts sidelobe features. This distortion is appreciable outside $|\mathrm{TelX}|\gtrsim0.25$ but negligible in the main beam where fits were performed. As a result, our determination of the centroid location and FWHM of the primary beam are not significantly distorted in the orthographic projection.

\begin{table}[ht]
\centering
\begin{tabular}{|c|c|c|c|}
\hline
Data Set & Solar & Holography (transit) & Drone \\\hline
Frequency Bins & 1024 & 1024 & 1024 \\
Feeds & 1 & 256 & 256 \\
Polarizations & 2 & 2 & 2 \\
TelX & -1.0 -- +1.0 & -0.5 -- +0.5 & -0.25 -- +0.25 \\
TelY & -1.0 -- 0 & (-0.88, -0.70, \textbf{-0.59},  & -0.62 -- -0.49 \\
& & -0.46, -0.15, 0.16) & \\ \hline
\end{tabular} \vspace{0.2cm}
\caption{\label{tab:surveyparams} Survey coverage and parameter space of each data set. Virgo A for the holographic data set is in bold. More details are given in the text.}
\end{table}

\subsection{Main Beam Fitting}
\label{sec:fitting}
\begin{figure}
\centerline{\includegraphics[width=3.5in]{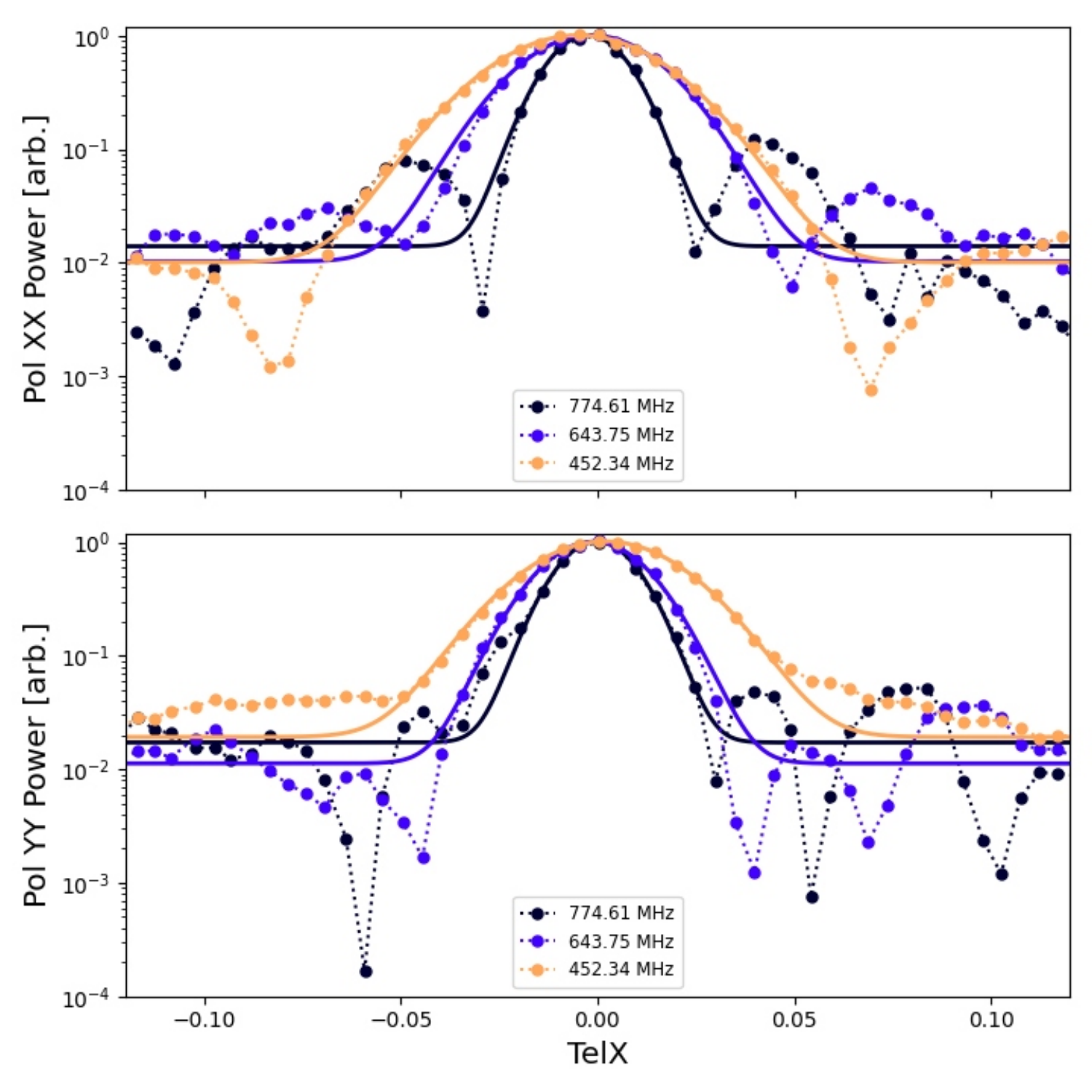}}
\caption{CHIME data (dotted lines) acquired during the drone flight. The power measured by feed 190 (corresponding to TelY coordinate -0.59 and declination 13.15$^{\circ}$) is shown at three frequencies in both polarizations (XX top, YY bottom). The power has been normalized by the amplitude of the best-fit 1d Gaussian from Equation~\ref{eqn:1dGauss} (solid line). The beam narrows with increasing frequency, as expected, and frequency-dependent sidelobe structure beyond the main beam shape is also apparent in the data. \label{Gauss1D}}
\end{figure}

We perform an unweighted least-squares fit of a 1d Gaussian function to the flight data in TelX to determine four parameters, as shown in Figure~\ref{Gauss1D}. For a single feed and frequency, the power $P$ as a function of TelX received as a source transits at a fixed TelY is:

\begin{equation}
    P(\mathrm{TelX}) = A e^{-\frac{(\mathrm{TelX}-\mathrm{TelX_0})^2}{2\sigma^2}} + B
    \label{eqn:1dGauss}
\end{equation}
where $A$ is the amplitude, $B$ is the background level, $\mathrm{TelX_0}$ is the centroid location in TelX, and $2.355*\sigma$ is the FWHM. Each combination of feed index, frequency, and polarization is analyzed separately for the drone and holography datasets for CHIME cylinder C.

\subsection{Altitude Estimation}
\label{sec:altitude}
\begin{figure}
\centerline{\includegraphics[width=3.5in]{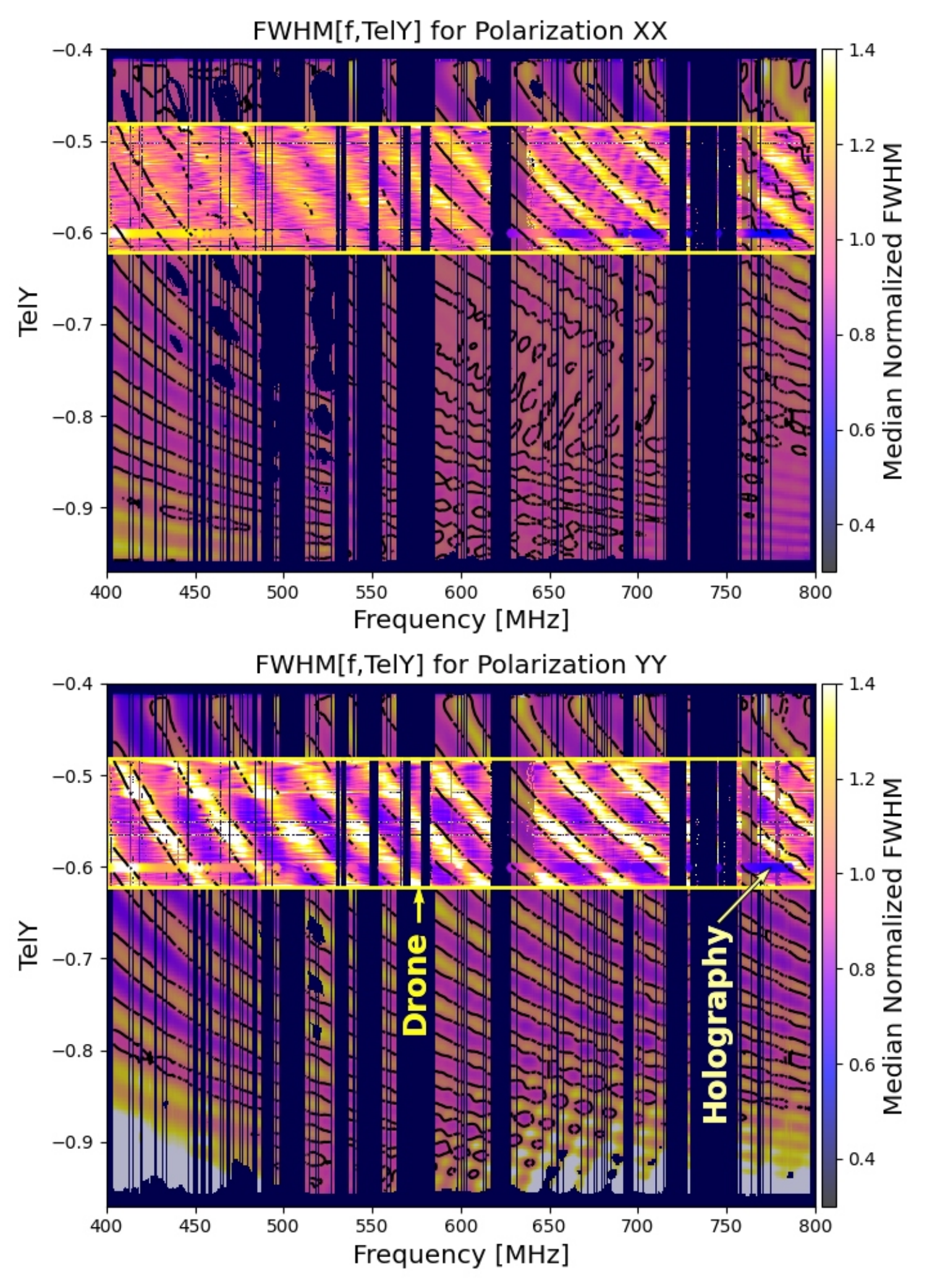}}
\caption{Best-fit FWHM as a function of frequency and altitude for the XX (top) and YY (bottom) polarizations. In each panel, the Solar data spans the coordinate space, the Drone data is overlaid for TelY within the range of (-0.62,-0.48) and indicated by the yellow box, and the Holography measurements from Virgo A are plotted as a band at TelY$\sim-0.6$. The colors are normalized to the median value within each frequency bin, and a contour at the median value has been traced over the solar data to enhance the appearance of the 30MHz ripple feature. The evolution of the FWHM parameter follows the same contour for all three data sets, and shows the characteristic 30MHz frequency ripple moving predictably as the altitude/declination increases. We examine FWHM as a function of frequency for a single feed in Figure \ref{CHIMEFWHM}, which is equivalent to a 1d horizontal slice through this plot at TelY $\approx -0.6$. \label{CHIMEFWHMfY}}
\end{figure}

In Figure \ref{CHIMEFWHMfY} we present the FWHM in each polarization as a function of frequency and TelY for all three data sources. The FWHM parameter for solar and drone data at each frequency has been normalized by the median value across the TelY axis. The solar data spans the shown coordinate space, and is present in the background of each plot in the figure.  The holography data (normalized with the median across the frequency axis instead) is overlaid at TelY=-0.6, neatly fitting within the ripple pattern present in the solar FWHM results. The FWHM results for the drone data, after accounting for the altitude estimation described in this section, are overlaid at the associated TelY $\in [-.62,-.49]$, and fall within the median-valued contour for the solar data at the associated declination.

Due to well-known multi-path effects within CHIME, the FWHM is a strong function of both TelY and frequency \cite{dallassolar}. This can be seen in the background of Figure \ref{CHIMEFWHMfY}, which shows FWHM values derived from solar data as a function of frequency and TelY. The solar data clearly has curving stripes with a 30MHz frequency ripple (associated with the distance between the telescope vertex and focus) and the peaks of the ripples shift with location along the North-South axis (TelY). We can include the drone FWHM on these plots, and determine if our mapping between feed number and solar TelY values is correct. Because we are using the ripple pattern to assess the mapping between TelY and feed, we normalize by the median FWHM value in each frequency bin. Because the actual FWHM values do not always agree (see more detail in Section \ref{sec:results}), this allows us to match using the contrast between the peaks and troughs, instead of the values themselves. While exploring the evolution of the FWHM parameter in frequency and TelY space we were able to determine the drone flight altitude with only a mild prior on the drone altitude from the drone sensors. This best-fit altitude differed from the altitude reported by the RTK system by a constant offset of $\sim5$\,m.

Using an altitude value of 307.6\,m above ground (311.6\,m above the image plane) from the drone's internal altimeter and RTK sensors, the drone FWHM results appeared offset by TelY $\approx-0.005$ relative to the solar FWHM pattern. We estimated that an $\sim5$\,m offset in flight altitude would account for this shift in TelY, suggesting the flight altitude was instead 312.6\,m (316.6\,m above the image plane).  To find a more precise value of this offset, we use a Pearson R (PR)\footnote{\href{https://docs.scipy.org/doc/scipy/reference/generated/scipy.stats.pearsonr.html}{Pearson R}} correlation coefficient computation as follows. For a given feed, the drone position and altitude define an angle which we convert to TelY, as described in Section~\ref{sec:mapproj}. Plotting data from all feeds produces a swath of FWHM values as a function of TelY and frequency, as shown in Figure~\ref{CHIMEFWHMfY}. We can vary the altitude of the drone, which shifts the swath up or down in TelY, and compare the resulting FWHM values to the solar data in the same swath. If the swath in TelY for a given altitude is incorrect, the peaks and troughs in the drone data won't align with the peaks and troughs in the solar data. We vary the drone altitude and compute the PR correlation coefficient between the drone and solar FWHM values. The altitude value which maximizes the PR coefficient provides a more precise estimate of the flight altitude than is given by the drone sensors. This is computed for each feed independently and we do not impose any requirements (e.g., that the FWHM pattern of all feeds taken together must match the solar data with a given geometric model).

\begin{figure}
\centerline{\includegraphics[width=3.5in]{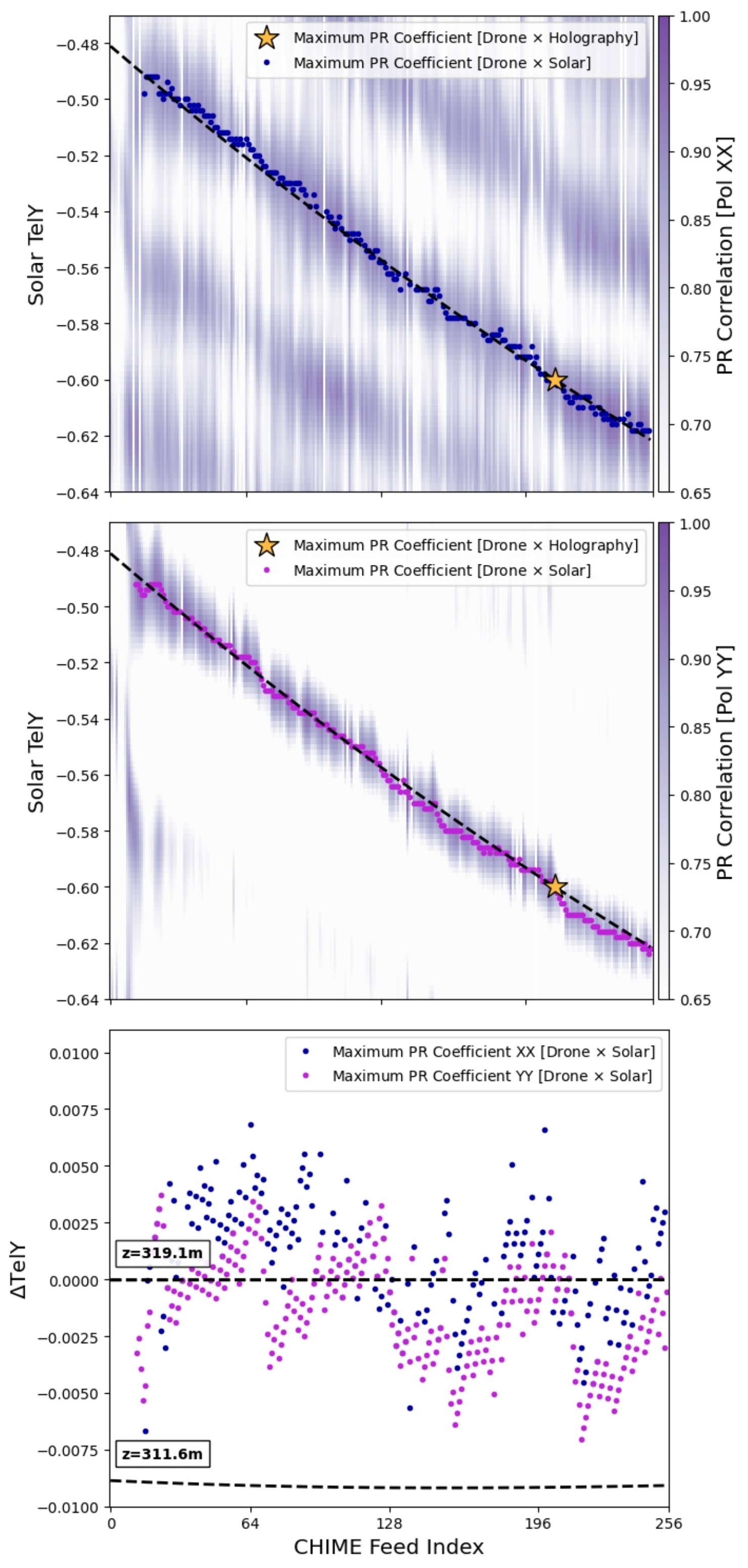}}
\caption{Pearson-R (PR) correlation coefficients calculated between drone data and solar data for each polarization (top XX, middle YY) of the 256 CHIME feeds mapped by the drone. The maximum value in each column (color dots) associates a CHIME feed with a TelY bin in the solar data. Lines indicate the best-fit altitude (319.1\,m). The maximum PR coefficient between the drone and Virgo A holography data occurs at feed 210 (yellow star) and falls along the TelY and feed mapping obtained from correlations with solar data. The bottom panel shows the and maximum PR correlation coefficients (blue and purple dots) for both polarizations after subtracting the TelY and feed mapping prescribed by the geometric model. An additional line is shown for the altitude given by the drone sensors (311.6\,m) in the bottom panel. \label{PRcoordfits}}
\end{figure}

The PR coefficients are shown in the upper two panels of Figure~\ref{PRcoordfits}, associating each of the 256 feeds used in the drone measurement (x-axis) with a Solar TelY value (y-axis). The results are roughly consistent with a constant altitude offset, a flight height of 315.1\,m above the ground (319.1\,m above the image plane), indicated by the black dashed lines in Figure \ref{PRcoordfits}. The two upper panels in Figure \ref{PRcoordfits} also show that the PR coefficients give a unique mapping between TelY and feed location for the YY polarization; however for the XX polarization, additional offsets also correlate strongly, although still has a clear maximum at the same location as the YY solution. As we show below, we suspect this is because the FWHMs for the XX polarization have more RFI-flagged frequencies and additional substructure in the FWHM pattern beyond the 30\,MHz ripple. This analysis suggests the magnitude of the global altitude offset is +7.5\,m relative to the altitude reported by the drone. 

Similarly, we compute the PR correlation between the drone and holography FWHM values across all frequencies for each feed. The maximum PR value thus represents which feed best matches the holography data from VirgoA (at TelY=-0.6). We find that the maximum value occurs at feed 210, which is indicated with a star in Figure \ref{PRcoordfits}. As expected, this independent determination of the feed to TelY relationship aligns with the mapping obtained from the drone and solar PR comparisons.

The lower panel of Figure \ref{PRcoordfits} shows the maximum PR points after subtracting the TelY per feed mapping prescribed by the geometric model. After this subtraction, a clear linear trend remains between TelY and chime feed, reminiscent of the centroid offsets per feed (see Section~\ref{sec:fitting}) which were not accounted for in this analysis. In addition, a residual sawtooth pattern is present spaced roughly every 64 feeds. Because solar data combines measurements from all cylinders while the drone data is generated from a single cylinder we would expect differences between the two data sets due to cylinder-dependent effects.

This offset could arise from inaccurate altimeter or RTK readings, a failure to account for height differences between the CHIME Center position provided by surveys and the terrain elevation directly below the drone transit trajectory, or differences between the terrain model used by the drone autopilot software and the site surveying. Additionally, because our drone platform produces several altitude estimation variables in each flight data file, which can be discrepant by up to 10\,m, it is not clear which variable (e.g. RTK, barometric altimeter, inertial measurement unit) is ultimately the most accurate. Further investigation is ongoing to assess the height accuracy of the drone sensors.


\section{RESULTS}
\label{sec:results}
\subsection{Centroid and FWHM Measurements}
\begin{figure}
\centerline{\includegraphics[width=3.5in]{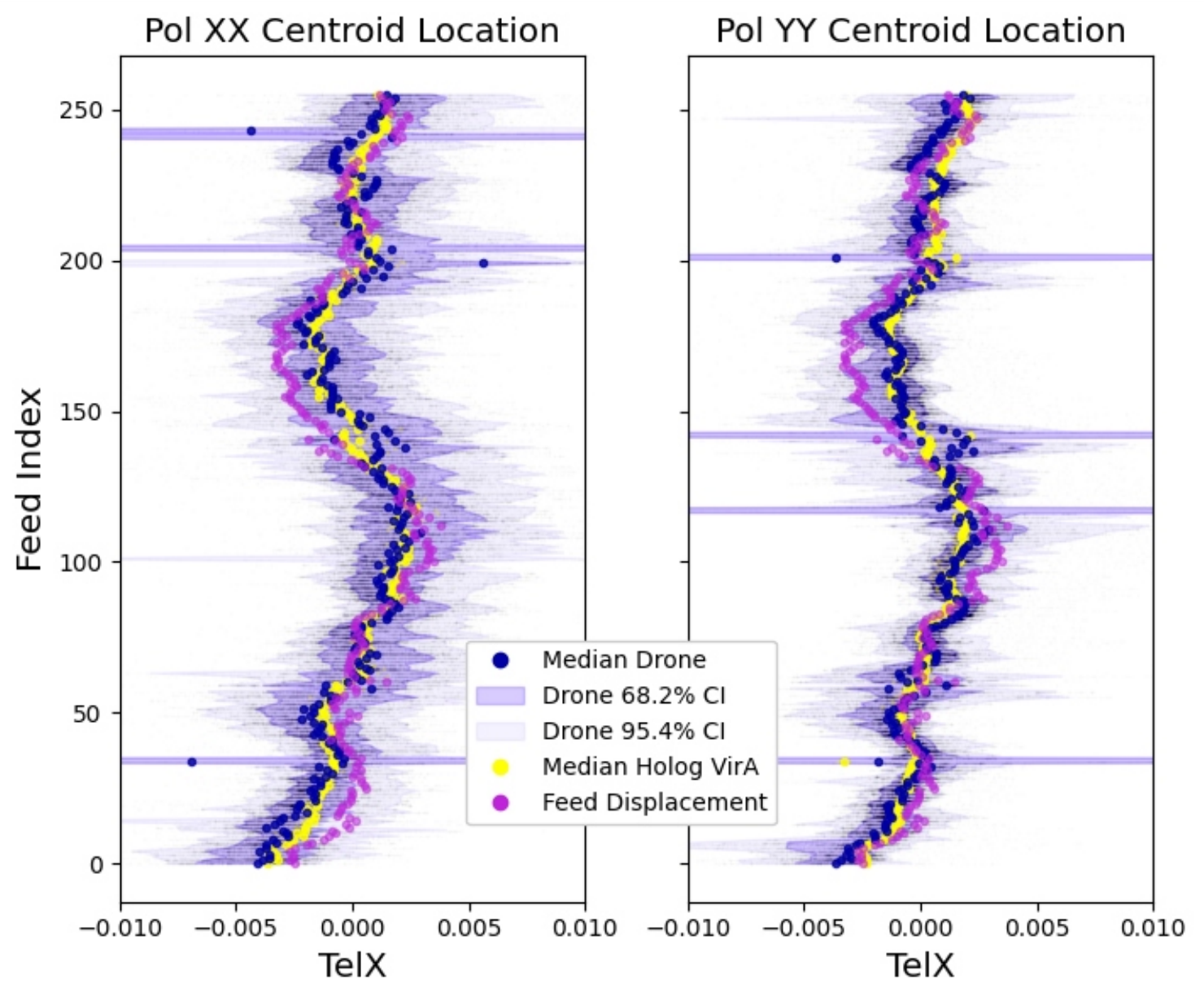}}
\caption{Best-fit centroid position from Drone and Holography data sets as a function of TelX for both XX (left panel) and YY (right panel) polarizations. Median centroid positions for all 256 feeds (median across all 1024 frequencies) from the drone data set are shown (navy dots). Shaded navy contours indicate 68.2\% and 95.4\% confidence intervals. Centroid positions from holography of Virgo A are similarly indicated (yellow dots), and can be compared with similar plots in \cite{CHIMEoverview,Redaholography}. These centroid measurements are consistent with photogrammetric measurements of the feed displacement presented in a doctoral thesis\cite{Dengthesis} (large purple dots). All data sets have been mean-subtracted such that absolute offsets are not included. \label{CHIMEcent}}
\end{figure}

From the best-fit 1d Gaussian (Equation \ref{eqn:1dGauss}), we obtain centroid and FWHM parameters as described in \ref{sec:fitting}. Both polarizations are measured independently in subsequent transits at the same zenith angle. Figure~\ref{CHIMEcent} shows the best-fit centroid values from the drone and holography data and compares them to photographic measurements performed along the CHIME feedline to measure feed displacement along the focal line \cite{Dengthesis} Because the solar data does not include any feed-level data, it is not represented in Figure~\ref{CHIMEcent}. Although the drone measurements are in the near-field, the centroid values and trend along the focal line from the drone agree with the two far-field measurements and follow the feed displacements. The centroid trend is known, and was previously published in \cite{CHIMEoverview}.

\begin{figure}
\centerline{\includegraphics[width=3.5in]{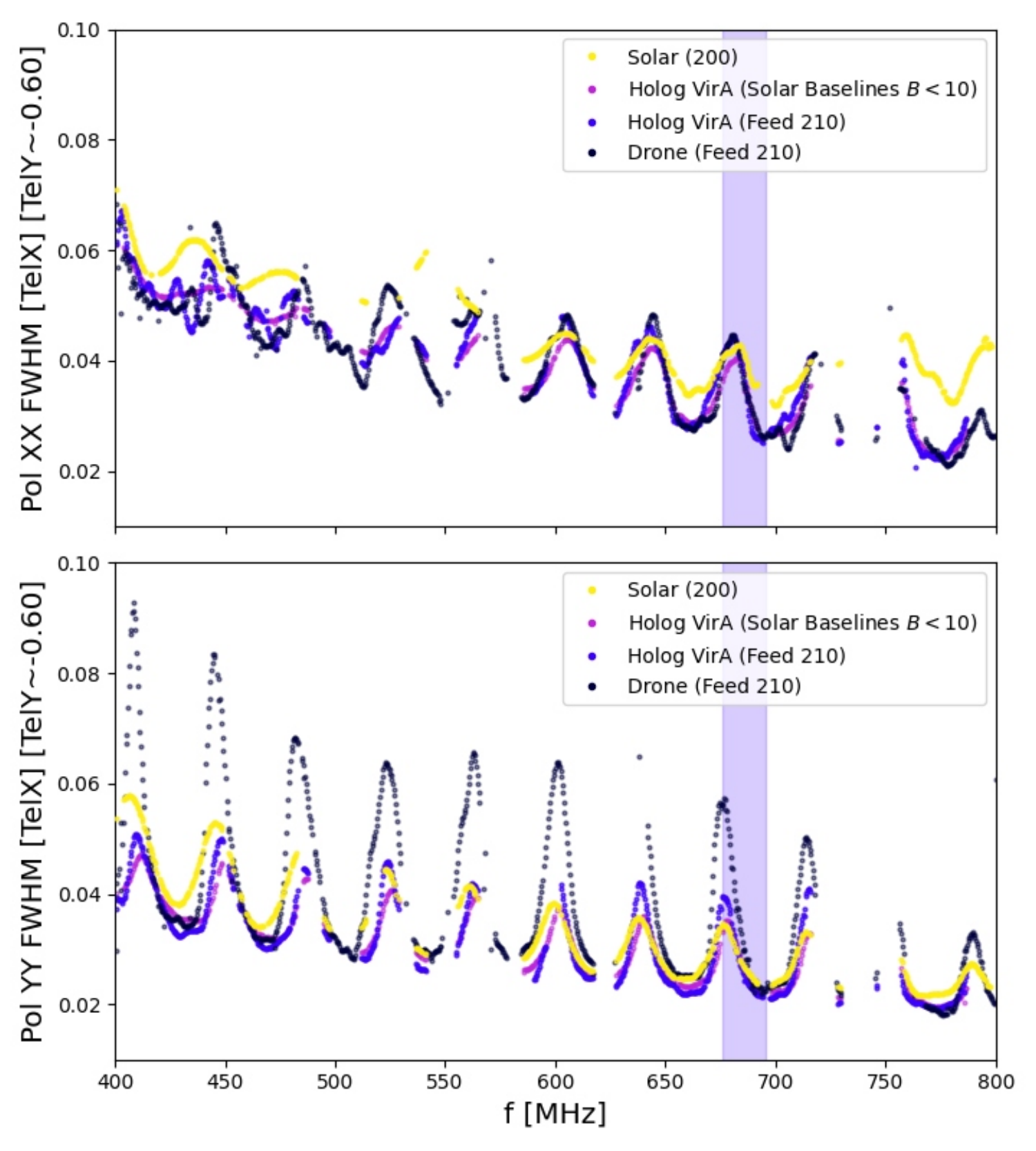}}
\caption{Best-fit FWHM from Drone, Holography (both for a single feed and an average of baselines $\le10m$ to form a data set equivalent to the solar data), and Solar data sets as a function of TelX. These curves can be directly compared because they all originate from a TelY of -0.6, where all three data sets overlap (see Figure \ref{overlap}). The holography and drone measurements are most comparable, because they can be separated into individual feeds. Conversely, the solar measurements are comprised of an average of all intercylinder baselines under 10m. To examine the divergence of the YY Polarization drone FWHM curve from the solar and holography datasets, we investigate transit data in the frequency band (676.2-695.7\,MHz) within the purple region in Figure~\ref{CHIMEYYdeviation}. To examine trends in FWHM as a function of both frequency and TelY, refer to Figure \ref{CHIMEFWHMfY}.
\label{CHIMEFWHM}}
\end{figure}

The best-fit FWHM parameter at TelY $\approx -0.6$ as a function of frequency is shown in Figure \ref{CHIMEFWHM} for three data sets: the solar data (formed from combining baselines less than 10\,m), the drone and holography data from a single feed, and the holography data with feeds within 10\,m combined to form a data set equivalent to the solar data. The characteristic 30\,MHz ripple feature is aligned for both polarizations at all frequencies, as one would expect from Figure~\ref{CHIMEFWHMfY}. The amplitude of the ripple feature is generally comparable between all data sets for the XX polarization, though the solar FWHM is wider than the holography and drone data sets throughout the frequency band. For the YY polarization, we find that the combined-feed holography data matches its solar counterpart very well, while the ripple amplitude is slightly higher for the single-feed holography data compared to the combined-feed holography data. This suggests that feed-to-feed variations may influence the amplitude of the ripple feature. We also find the peak FWHM amplitude found in the drone data is typically significantly higher than the other data sets for YY polarization, which prompted further investigation. 

\begin{figure}
\centerline{\includegraphics[width=3.5in]{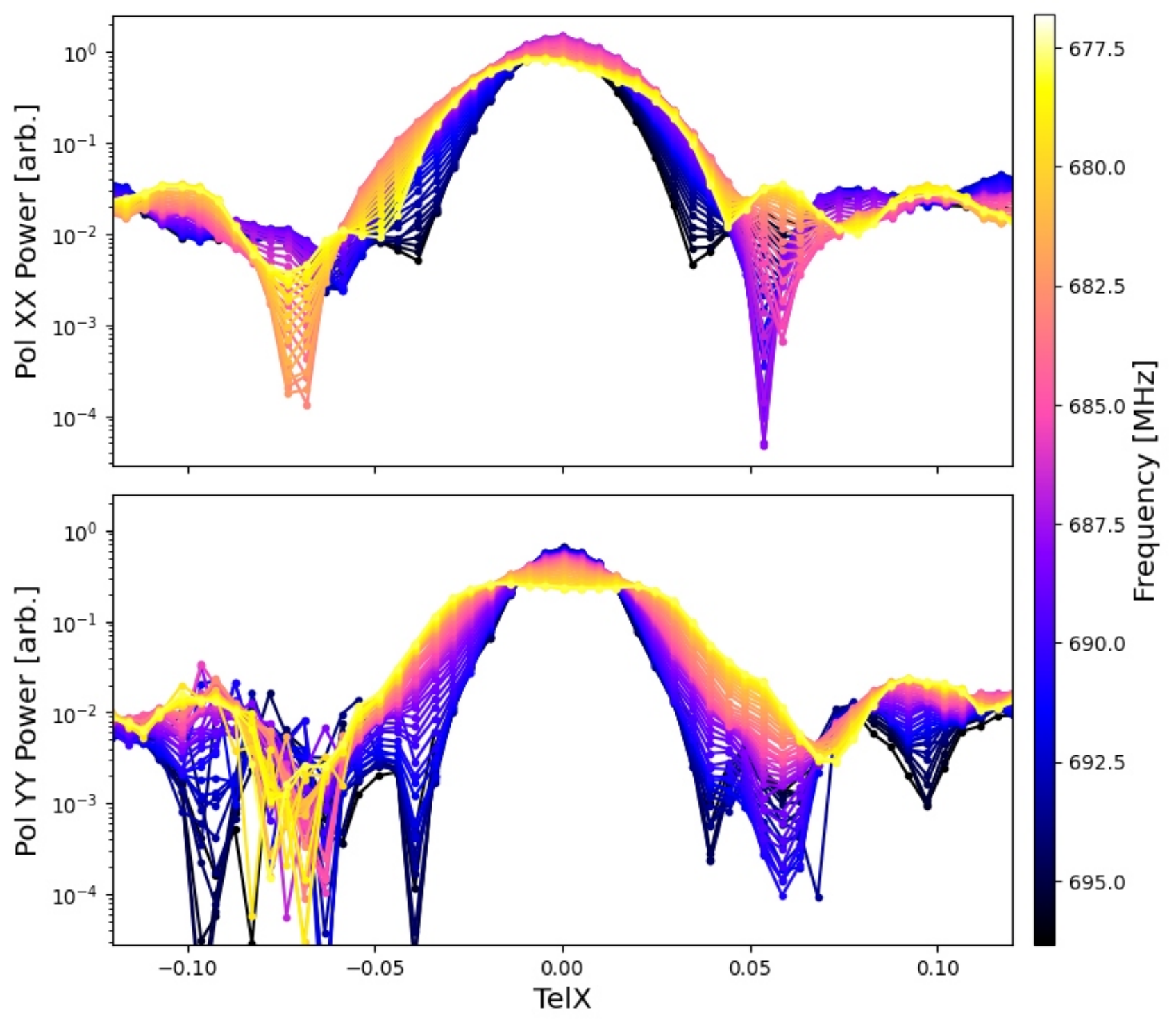}}
\caption{First eigenmode data from autocorrelation measurements of the drone transiting at TelY $\approx -0.6$ (feed 210) across CHIME cylinder C. The data shown are not normalized or corrected with the best-fit centroid. The frequency band shown here is reversed from the frequency axis of Figure~\ref{CHIMEFWHM}. The colored curves thus evolve along the 30\,MHz ripple feature in the FWHM$(f)$ curve starting from a minima at 695.7\,MHz (black), where the FWHM$(f)$ values are in agreement with the solar and holography datasets, to a maxima at 676.2\,MHz (yellow), where the divergence is highest. In the logarithmic scale shown in Figure~\ref{CHIMEYYdeviation} the amplitude appears compressed, but in linear units the amplitude does not appear compressed.
\label{CHIMEYYdeviation}}
\end{figure}

Figure~\ref{CHIMEYYdeviation} contains drone transit measurements for frequencies within 676.2-695.7\,MHz. This frequency band follows the evolution of the 30\,MHz ripple feature in the FWHM$(f)$ curve from a maximum to a minimum, indicated by purple region in Figure \ref{CHIMEFWHM}. In both polarizations, we expect to find a well defined main beam at TelX=0.0 that ends at a first null around $\pm0.04\le\textrm{TelX}\le\pm0.05$). Wherever the 30\,MHz ripple feature reaches a local minimum, (e.g. $f=695$\,MHz plotted in black in Figure~\ref{CHIMEYYdeviation},) the main beam and sidelobes are discrete and the best-fit Gaussian FWHM values agree with the solar and holography datasets as expected for both the XX (top panel) and YY (bottom panel) polarizations. Conversely, wherever the 30\,MHz ripple reaches a local maximum, the primary beam and first sidelobe are not separated by a null in the drone transit data. Instead, the main beam and first sidelobe are combined and the Gaussian fit finds a FWHM that is wider than expected. Examples can be found in both the XX polarization (e.g. $f=685.0$\,MHz plotted in pink in the top panel of Figure~\ref{CHIMEYYdeviation}) and YY polarization (e.g. $f=677.5$\,MHz plotted in yellow in the bottom panel of Figure~\ref{CHIMEYYdeviation}).

To demonstrate this behavior, we compare both polarizations at $f=677.5$\,MHz (plotted in yellow). The XX polarization data clearly recovers the main beam and two well-defined sidelobes. The YY polarization data only recovers the second sidelobe (centered at $\textrm{TelX}\approx\pm0.1$) while the first sidelobe (centered at $\textrm{TelX}\approx\pm0.05$ in XX) been been incorporated into the main beam. This behavior evolves smoothly in both polarizations as we progress along the gradient of plotted frequencies. The XX polarization recovers the expected first null on either side of the main beam at the highest and lowest plotted frequencies, but intermediate frequencies show the first sidelobe joining the main beam (particularly for $\textrm{TelX}<0$). The evolution for the YY polarization is similar but not identical---as the plotted frequency decreases, the sharpness of the null diminishes, eroding the distinction between the main beam and the first sidelobe. 

The cause of the differences between the celestial source data and the drone data sets (most prominently in the YY polarization) are not entirely understood. One possibility is that this could be due to near-field effects. Where the main beam is widest, near-field effects may combine the main beam and first sidelobe, failing to recover the first order null that exists in the far-field beam, resulting in the behavior described above. It is not clear why this would impact one polarization more than the other, although the two polarizations do have different illumination patterns. Similarly, the amplitude of this discrepancy is not common across all feeds: feed numbers between 80-90 and 220-230 (see Figure \ref{CHIMEray}) are in agreement with the FWHM amplitudes from other data sets. The discrepancy therefore can't be entirely explained by sampling in the radiating near-field, because feeds 80-90 are $\sim25\,$m closer to the drone than feed the plotted feed (210). Additional flights would be helpful for investigating this discrepancy, but are beyond the scope of this paper.  


\subsection{Trends with Frequency and TelY}
The solar and drone data can be compared in the TelX vs. frequency and TelX vs. TelY subspaces to further investigate beam patterns. The drone data is taken from the first eigenmode (see Section~\ref{sec:methodology}.\ref{sec:config}) to increase the signal to noise. As previously described, the solar data is an average of all intercylinder baselines $\le10m$. All data has been corrected for the frequency-dependent centroid deviation measured in the Gaussian fits. 

\begin{figure}
\centerline{\includegraphics[width=3.5in]{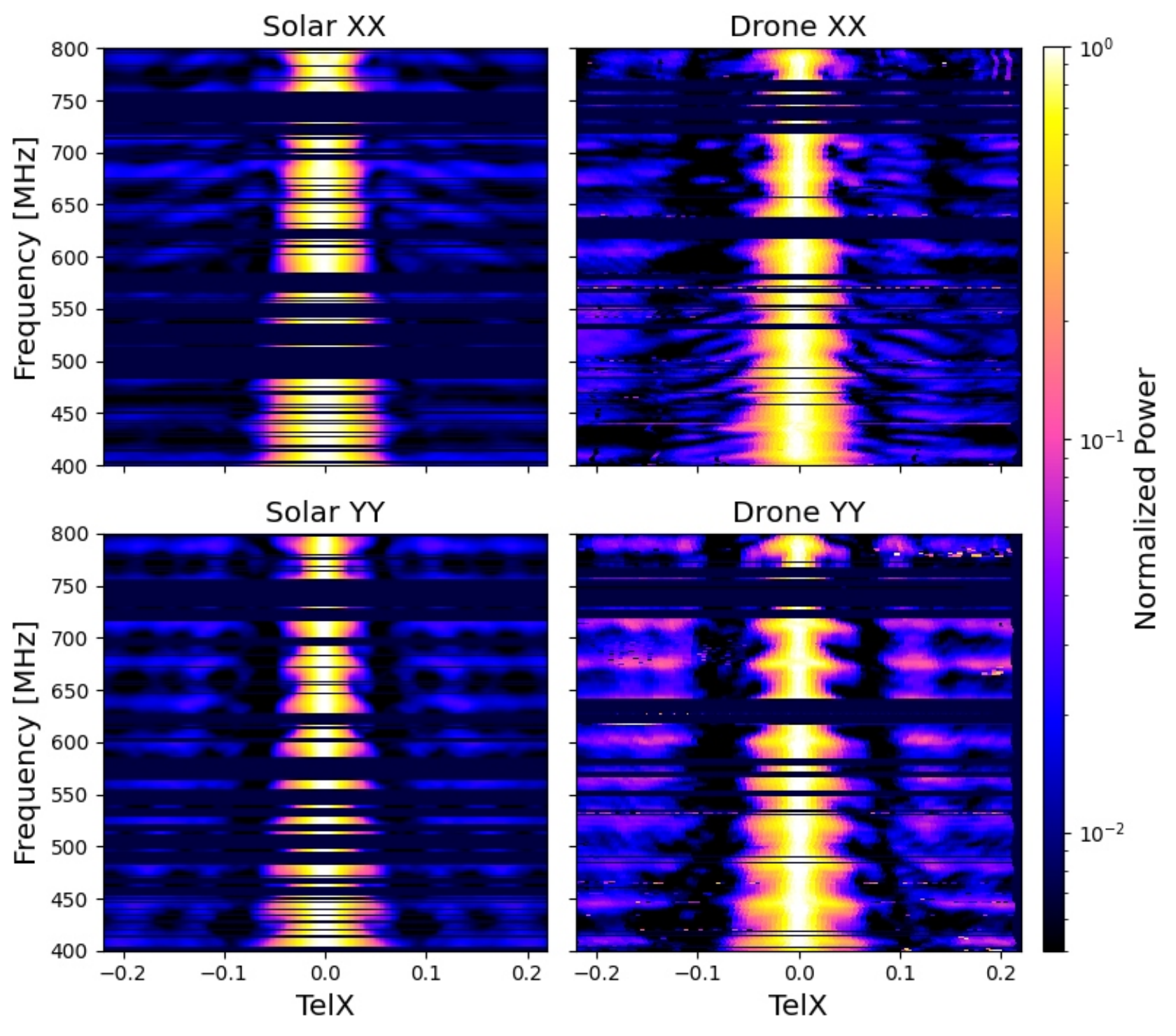}}
\caption{Waterfall plots of telescope data obtained during Drone measurements (Right Column, single feed [210] in either polarization) and Solar measurements (Left Column, average of intercylinder baselines $\le10m$ in either polarization). Similar widths and features are present in both data sets, and the beamwidths and sidelobe features are comparable. Each frequency is normalized independently to a Gaussian fit peak value to remove the uncalibrated frequency dependence of the drone transmitter.  \label{CHIMEfX}}
\end{figure}

First, we compare the drone and solar data as a function of TelX and frequency in Figure \ref{CHIMEfX}. For this comparison we select a single TelY index for the solar transit data and a single CHIME feed index for the drone data (both corresponding to $TelY=-0.600$, feed 210). The amplitude of the drone and solar data have been normalized in each frequency with the amplitude of the associated best-fit gaussian. These waterfall plots show similar frequency dependent variations in width in the main beam and sidelobe amplitudes and positions. Specifically, the width decreases with increasing frequency, and is modulated by a 30\,MHz ripple aligned between the two data sets (as would be expected from Figure~\ref{CHIMEFWHM}). Similarly, the XX polarization has a wider overall beam than the YY polarization, also as expected from the CHIME feed illumination pattern\cite{Deng_clover:2017, CHIMEoverview}. The drone data has a more pronounced ripple in the XX polarization, whereas the ripple is attenuated in the solar data due to averaging over many baselines (supported by the lower ripple amplitude as described in Section~\ref{sec:methodology}.\ref{sec:fitting}). 

\begin{figure}
\centerline{\includegraphics[width=3.5in]{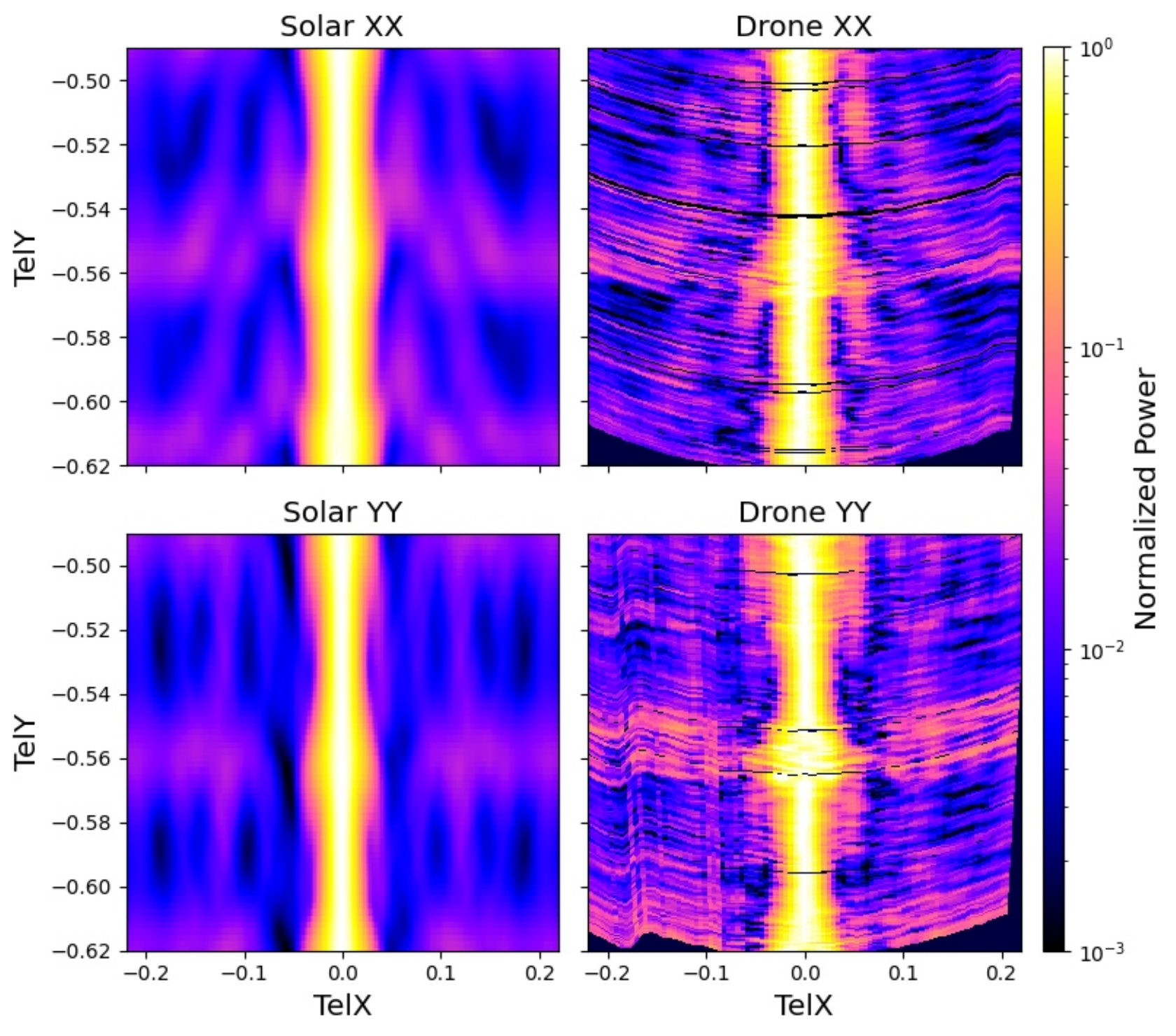}}
\caption{Waterfall plots of telescope data obtained during Solar measurements (average of intercylinder baselines $\le10$ for a single frequency bin (center 690MHz) in XX (Left Top) and YY (Left Bottom) polarizations) and Drone measurements (all feeds for a single frequency bin (center 690\,MHz) in XX (Right Top) and YY (Right Bottom) polarizations). \label{CHIMEXY}}
\end{figure}

Second, we compare the drone and solar data as a function of TelX and TelY in Figure \ref{CHIMEXY}. In this figure, a single frequency bin (centered around 690.62\,MHz) is selected for the drone and solar data. The TelY axis for the drone data is constructed from each separate feed within cylinder C from the drone transit using the mapping computed in Section \ref{sec:methodology}.\ref{sec:altitude}. The drone data and solar data are normalized by the amplitude of the best-fit Gaussian along the TelY axis. 

For both polarizations, the main beam and sidelobes have similar TelY dependence in both the drone and solar datasets. 
A notable difference between these datasets is the high signal-to-noise and continuity present in the solar data from coadding in baseline and time. Conversely, each of the TelY values in the drone data comes from the same 3.4\,min drone flight, and does not benefit from coadding or averaging. However, signal-to-noise improvements are possible with modest increases in flight time. 

Comparing the solar and drone data for the YY polarization (bottom panels of Figure \ref{CHIMEXY}), the following similarities are apparent:
The main beam expands and shrinks at the same values of TelY (expands at TelY=-0.56, two clear `waists' at TelY $\sim$-0.53 and -0.6). The sidelobe levels are higher where the beam widens (TelY=-0.56) and lower at TelY values where the main beam shrinks to a waist. The third order sidelobe is maximal at TelX=$\pm0.125$ for the plotted range of TelY, appearing as a continuous pink line. 

Turning to the XX polarization (top panels of Figure \ref{CHIMEXY}), we observe the same similarities across the drone and solar data. The main beam broadens at TelY=-0.55, and shrinks to a waist at TelY=-0.52 and TelY=-0.59 in both data sets. Unlike in polarization YY, the sidelobe levels for polarization XX evolve in TelX and TelY simultaneously---shown by symmetric sloped contours that extend away from waist in the main beam (e.g. starting at TelY=-0.59, and expanding in TelX as TelY decreases to -0.62.)

\section{CONCLUSION AND FUTURE DIRECTIONS}
\label{sec:conc}
The measurements presented in this paper represent the successful development of a correlator data acquisition mode for drone-based beam calibration and demonstrate the first use of a drone as a reference calibrator for a large cylindrical radio interferometer. The comparisons between the drone, holography, and solar measurements indicate that the FWHM and centroid position are consistent across all three data sets. Moreover, the evolution of these parameters as a function of declination, frequency, and feed index are shown to behave similarly in all three data sets, despite differences in measurement technique. This shows that drone beam measurements are possible, even at the relatively slow data cadence of CHIME, by using its pulsar gating mode. The analysis also demonstrates the complexities of measuring a cylindrical array with a drone and highlighted the areas for improvement, in particular a higher precision constraint on drone altitude is essential for this mapping and designing flight trajectories which can more easily be mapped to comparison data sets.

During the same maintenance window, our group conducted nine other drone beam mapping test flights which are prime targets for future analysis. We performed similar East-to-West scans at 45$^{\circ}$ and 55$^{\circ}$ zenith angle, North-to-South transits along the focal line of Cylinder C, and gridded flights over the center of the array. In addition, test flights were performed where the drone rotated through 180$^{\circ}$ of heading (and thus, polarization) at several locations above the array to test the polarization response of the feeds as a function of angle.

The RFI from the drone contributes unwanted signal in otherwise low-background sites. We are investigating using a new drone, with higher frequency communications (shifted far beyond the restricted range at DRAO), which also has longer flight times, and improved time and location accuracy.

Now that the correlator firmware and pulsar timing parameter files have been generated, we can also return to CHIME to conduct additional drone flights targeting regions of interest within the beam, including two-dimensional polarized properties of the beam maps. Some limitations noted in this paper are important to address in these future flights. First, neglecting the transmitter beam is currently a large sytematic error for angles beyond the main beam, and will need to be incorporated and removed for future work. Along these lines, making measurements equivalent to the more standard Ludwig-III polarization definition are important for future work.



\section*{ACKNOWLEDGMENT}
The authors would like to acknowledge contributions from the DRAO staff and faculty.

We thank the Dominion Radio Astrophysical Observatory, operated by the National
Research Council Canada, for gracious hospitality and expertise. The DRAO is
situated on the traditional, ancestral, and unceded territory of the Syilx
Okanagan people. We are fortunate to live and work on these lands.

We extend special thanks to Tim Robishaw, Alex Hill, Tom Landecker, Benoit Robert, Kory Phillips, Mandana Amiri, Niko Milutinovic, Michael Rupen, Richard Hellyer, Rick Smegal, and Luna Zagorac for assistance during flights at the DRAO, including coordination across the DRAO telescopes and facilities.

CHIME is funded by grants from the Canada Foundation for Innovation (CFI) 2012
Leading Edge Fund (Project 31170), the CFI 2015 Innovation Fund (Project 33213),
and by contributions from the provinces of British Columbia, Québec, and
Ontario. Long-term data storage and computational support for analysis is
provided by WestGrid, SciNet and the Digital Research Alliance of Canada, and we
thank their staff for flexibility and technical expertise that has been
essential to this work.

Additional support was provided by the University of British Columbia, McGill
University, and the University of Toronto. CHIME also benefits from NSERC
Discovery Grants to several researchers, funding from the Canadian Institute for
Advanced Research (CIFAR), from Canada Research Chairs, from the FRQNT Centre de
Recherche en Astrophysique du Québec (CRAQ) and from the Dunlap Institute for
Astronomy and Astrophysics at the University of Toronto, which is funded through
an endowment established by the David Dunlap family. This material is partly based on work supported by the Perimeter Institute for Theoretical Physics, which in turn
is supported by the Government of Canada through Industry Canada and by the
Province of Ontario through the Ministry of Research and Innovation. L.B.N and A.R. acknowledge the support of the National Science Foundation award number 2006911; L.B.N and W.T. acknowledge the support of the NSF award number 1751763; A.O. is partly supported by the Dunlap Institute at the University of Toronto; and K.M.B. is supported by NSF grant 2018490. 




\bibliographystyle{unsrt} 
\bibliography{references}

\end{document}